\documentstyle[12pt,epsfig]{article}

\newcommand{\be}{\begin{equation}}
\newcommand{\ee}{\end{equation}}
\begin{document}
\topmargin 0pt
\oddsidemargin=-0.4truecm
\evensidemargin=-0.4truecm
\renewcommand{\thefootnote}{\fnsymbol{footnote}}
\newpage
\setcounter{page}{0}
\begin{titlepage}
\vspace*{-2.0cm}
\begin{flushright}
hep-ph/0511341
\end{flushright}
\vspace*{0.1cm}

\begin{center}
{\title\bf\Large{Solar Neutrinos: Spin Flavour Precession and LMA}}
\vspace{0.6cm}

\vspace{0.4cm}

{\large
\underline{Jo\~{a}o Pulido\footnote{E-mail: pulido@cftp.ist.utl.pt}}, 
Bhag C. Chauhan\footnote{On leave from Govt. Degree College, Karsog (H P)
India 171304. E-mail: chauhan@cftp.ist.utl.pt} \\
\vspace{0.15cm}
{  {\small \sl CENTRO DE F\'{I}SICA TE\'{O}RICA DAS PART\'{I}CULAS (CFTP) \\
 Departamento de F\'\i sica, Instituto Superior T\'ecnico \\
Av. Rovisco Pais, P-1049-001 Lisboa, Portugal}\\
}}
\vspace{0.25cm}
and \\
\vspace{0.25cm}
{\large R. S. Raghavan\footnote{E-mail: raghavan@vt.edu}} \\
{\it Department of Physics  \\ Virginia Polytechnic Institute and State University
(Virginia Tech) \\  Blacksburg  VA 24060 USA}
\end{center}
\vglue 0.6truecm

\begin{abstract}
The time dependence that appears to be hinted by the data
from the first 13 years of the solar neutrino Gallium experiments is viewed as
resulting from a partial conversion of active neutrinos to light sterile ones
through the resonant interaction between the magnetic moment of the neutrino and
a varying solar field. A summary of the model and its predictions are presented
for the forthcoming experiments Borexino and LENS.
\end{abstract}
\end{titlepage}

{\bf\Large 1.}  After LMA has been ascertained as the dominant solution to the
solar neutrino problem, the next step in solar neutrino experiments will be the
search for time variations of the active flux. Such an effect remains much
of an open question because so far most analyses have relied on time averaged
solar neutrino data. Moreover, its investigation is of extreme importance
because it may reveal the existence of electromagnetic properties of
neutrinos and provide a significant addition to our understanding of
solar dynamics.

Although the idea of a sizeable neutrino magnetic moment interacting with
the solar field \cite{Cisneros:1970nq} lay dormant for some time, it was
reintroduced in 1986 \cite{Okun:1986hi} to explain the claimed periodicity
of the Chlorine event rate, but the effect remains inconclusive. In summary
it works as follows: active neutrinos can be
converted to sterile ones owing to the interaction of $\mu_{\nu}$ with
$B_{\odot}$. At times of intense solar activity
\begin{equation}
Strong~B_{\odot}~\rightarrow~large~\mu_{\nu}B_{\odot}~\rightarrow~large~
conversion
\end{equation}
with little or no conversion otherwise. Hence a neutrino flux anticorrelated to
solar activity.

\begin{center}
\begin{tabular}{ccc} \\ \hline \hline
Period &  1991-97 & 1998-03 \\ \hline
SAGE \& Ga-GNO & $77.8\pm 5.0$ & $63.3\pm 3.6$ \\
Ga-GNO only & $77.5\pm 7.7$ & $62.9\pm 6.0$ \\
no. of suspots     & 52           & 100 \\ \hline
\end{tabular}
\end{center}
{\it{Table I - Event rates in solar neutrino units for Ga experiments in
two different periods of solar activity and number of sunspots in
the same periods.}}

In table I the neutrino event rate in Gallium experiments (SAGE
and Gallex-GNO \cite{Cattadori})
is shown as a function of time with the number of sunspots in the same period.
It is seen from this table that there is a 2.4$\sigma$ discrepancy in the
combined results over the two periods. Since the sunspot number is correlated
to solar activity, this is suggestive of an anticorrelation of the Ga event
rate with the 11-year solar cycle. No other experiments show such a variational
effect, therefore since Ga are the only ones with a significant contribution of
$pp,~^7 Be$ neutrinos ($\simeq$ 80\%), the time dependence of these fluxes becomes
an open possibility.

\vspace{0.3cm}

{\bf\Large 2.}  The bottom of the solar convective zone is a region where,
owing to the relatively strong rotation gradient, the solar field is expected
to have a maximum which can be as high as 300 kG and may be connected to
the magnetic surface activity. The typical density profile of the sun requires
the mass square difference associated to resonant conversion from
active to nonactive neutrinos to be of the order of $10^{-8}eV^2$
\cite{Chauhan:2004sf},\cite{Chauhan:2005pn}.
This excludes the solar and atmospheric mass square differences,
$\Delta m^{2}_{21}$ and $\Delta m^{2}_{32}$, and therefore conversion to
$\bar\nu_{\mu}$ or $\bar\nu_{\tau}$. Since only three neutrino families are
known to exist, we must introduce sterile neutrinos. We consider the simplest
possible assumption whereby they mix with active ones through the magnetic moment
only, so that in the vacuum
\begin{equation}
\left(\begin{array}{c}\nu_{s}\\ \nu_{e}\\ \nu_{x}\end{array}\right)=
\left(\begin{array}{ccc}1&0&0\\ 0&
c_{\theta}&s_{\theta}\\ 0&-s_{\theta}&
c_{\theta}\end{array}\right)\left(\begin{array}{c}\nu_{0}\\ \nu_{1}\\
\nu_{2}\end{array}\right)
\end{equation}
while in matter with a magnetic field
\begin{equation}
\cal{H}_{\rm {M}}=\left(\begin{array}{ccc}\frac{-\Delta m^2_{10}}{2E}&
\mu_{\nu}B&0 \\ \mu_{\nu}B& \frac{\Delta m^2_{21}}{2E}s^2_{\theta}+V_e&
\frac{\Delta m^2_{21}}{4E}s_{2\theta}\\ 0&\frac{\Delta m^2_{21}}{4E}s_{2\theta}&
\frac{\Delta m^2_{21}}{2E}c^2_{\theta}+V_{x}\end{array}\right)
\end{equation}
with standard notation \footnote{For details see ref. \cite{Chauhan:2004sf}.}.
The 'new' mass square difference $\Delta m^2_{10}=m^2_{1}
-m^2_{0}$ fixes the location of the active $\rightarrow$ sterile transition
for each neutrino energy.

\vspace{0.6cm}

\begin{figure}[h]
\setlength{\unitlength}{1cm}
\begin{center}
\hspace*{-1.6cm}
\epsfig{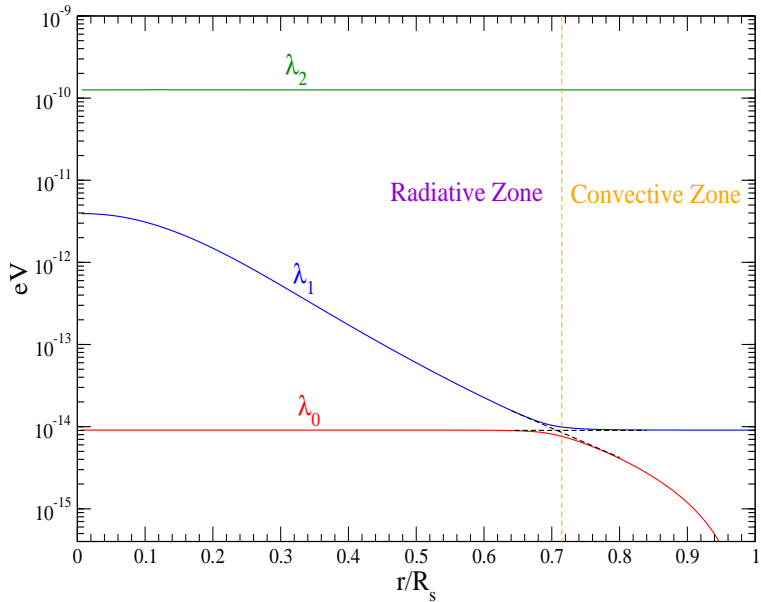}
\end{center}
\caption{ \it The evolution of the mass matter eigenvalues along the
solar neutrino trajectory.}
\end{figure}

For typical parameter values (peak field $B_0=2.2\times10^5~G,~\Delta m^2_{10}=
-6.0\times10^{-9}eV^2,~E=0.33MeV$) the eigenvalue evolution within the sun
is shown in fig.1. All pp neutrinos resonate around the bottom of the convective
zone in this case. It is seen that the strongly adiabatic LMA
transition, whose location is determined for each energy by $\Delta m^2_{21}$,
takes place in the solar core. The active $\rightarrow$ sterile transition,
whose adiabaticity depends on the magnetic field strength and its extension,
takes place around the upper radiative/lower convective zone (i. e. the
tachocline). The large order of magnitude difference between $\Delta m^2_{21}
~(7-8\times 10^{-5}eV^2)$ and $\Delta m^2_{10}~(-6\times 10^{-9}eV^2)$ ensures
the two resonances to be located far apart so they do not interfere.
In fig.1 the vanishing field case is denoted by the two dotted
lines crossing at the place where $\lambda_1$ and $\lambda_0$ nearly meet
(critical density).

The two field profiles chosen are peaked at the bottom of the convective zone and
can be seen in fig.1 of ref.\cite{Chauhan:2005pn}. If these are time dependent
(possibly connected to solar
activity), the effect provides a modulation of the pp neutrino flux mainly.
For larger $ \Delta m^2_{10}~(eV^2) $ ($ =-1.0\times 10^{-8}~eV^2$) the
spin flavour precession  resonance moves in the direction of the solar surface
and both $pp$ and $^7 Be$ fluxes become modulated ($pp+^7\!Be$ modulation).



\vspace{0.3cm}

{\bf\Large 3.}  Our knowledge of the solar neutrino spectrum is very limited.
$^8 B$, the best well known flux, accounts for just $10^{-4}$ of the total,
while $pp$ and $^7 Be$ accounting for more than 98\%, have only been observed
in the radiochemical Ga experiments in accumulation with all other fluxes.
In view of a hint from Ga experiments that low energy fluxes might be
time dependent, we need real time dedicated low energy experiments to
ascertain this possibility. We next consider the predictions for two of
the forthcoming ones: Borexino and LENS \cite{Chauhan:2004sf}.

\begin{center}
\bf{(I) Borexino}
\end{center}


This is a liquid scintillator detector installed in Gran Sasso Laboratory
using $\nu~e^{-}$ scattering, aiming at starting physics in late 2006. The
kinetic energy threshold of 250 keV and maximum of 664 keV ensures that this
experiment is directed mainly at $^7 Be$ neutrino flux ($E_{Be}=0.86MeV$).
The reduced rate for Borexino is, in standard notation
\begin{equation}
R_{Bor}=\frac{\int_{T_{min}}^{T_{max}}dT
\int_{E_{min}}^{E_{max}}dE\phi(E)[P_{ee}(E)\frac{d\sigma_{\nu_e}}{dT}
\!+\!\!(\!1\!-\!P_{ee}\!(E)\!)\frac{d\sigma_{\nu_x}}{dT}\!]}
{\int_{T_{min}}^{T_{max}}dT
\int_{E_{min}}^{E_{max}}dE\phi(E)\frac{d\sigma_{\nu_e}}{dT}}
\end{equation}
and is shown in fig.2 as a function of the peak field $B_0$. It is
seen that in the $pp+^7\!Be$ dominated modulation the rate decreases faster
for increasing $B_0$, thus exhibiting more sensitivity to solar activity
than in the $pp$ case: the more sensitive $^7 Be$ flux is to the peak field,
the more sensitive will the event rate be. As the Borexino collaboration expects
a combined error (statistical+systematic) of 10\% and 5\% after 1 and 3 years
run, both scenarios can be perceptible after 1 year.

\begin{figure}[h]
\setlength{\unitlength}{1cm}
\begin{center}
\hspace*{-1.6cm}
\epsfig{file=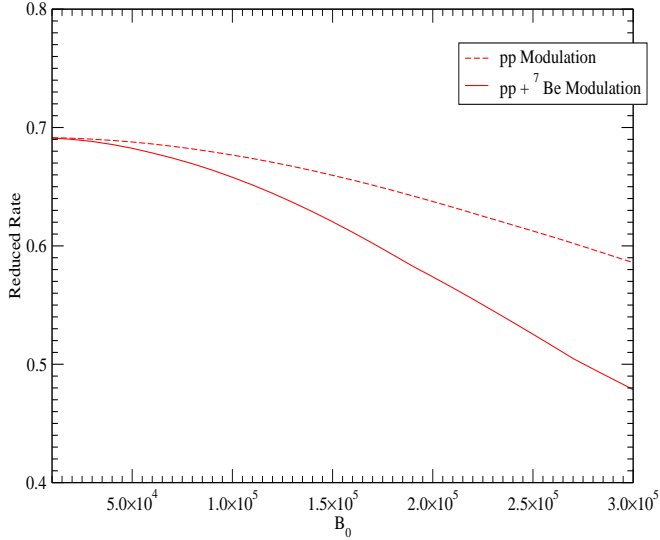,height=10.0cm,width=9.0cm,angle=270}
\end{center}
\caption{ \it Reduced Borexino event rate as a function of the peak field value.}
\end{figure}

\newpage
\begin{center}
\bf{(II) LENS (Low Energy Neutrino Spectroscopy)}
\end{center}

LENS is a real time detector in late stages of development measuring solar
neutrinos through the charged current reaction
\begin{equation}
\nu_e+^{115}In \rightarrow ^{115} Sn + e^{-}
\end{equation}
with the lowest threshold yet: Q=114 keV. The signal energy is directly and
uniquely related to the neutrino energy, hence a resolved spectrum of all low
energy components ($pp,~^7 Be,~pep,~CNO$) can be obtained that qualitatively
shows how the sun shines. The LENS event rate is
\begin{equation}
R_{_{LENS}}(E_e)=\int_Q^{E_{max}}P_{ee}(E)f(E^{'}_e,E_e)\phi(E)dE
\end{equation}
where $E^{'}_e$ is the physical electron energy, $E^{'}_e=E-Q$.
For $pp~+~^7 Be$ modulation ($\Delta m^{2}_{10}=-1.0\times 10^{-8}~eV^2$)
$R_{_{LENS}}(E_e)$ is shown in fig.3.
\begin{figure}[h]
\setlength{\unitlength}{1cm}
\begin{center}
\hspace*{-1.6cm}
\epsfig{file=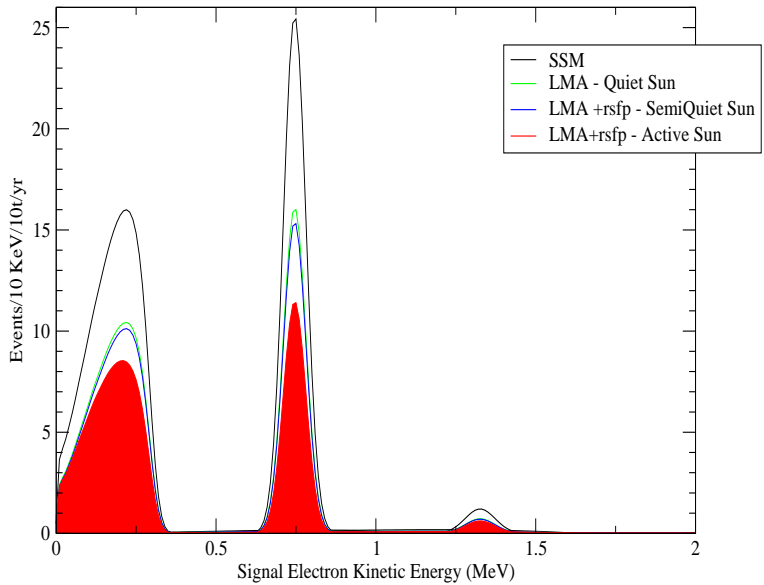,height=11.0cm,width=10.0cm,angle=270}
\end{center}
\caption{ \it LENS rate for $pp\!+^7\!\!Be$ modulation
and a peak field 250 kG for the lower line.}
\label{fig3}
\end{figure}

{\bf\Large 4.}  To conclude: low energy solar neutrino experiments should at
present be regarded as a major objective in the solar neutrino program, as time
modulation is hinted by the Gallium experiments. Running Borexino and LENS
during a significant fraction of a solar cycle will certainly test the
possible modulation effect in the low energy sector, thus providing evidence
for $\mu_{\nu}$ and new physics.


\begin{thebibliography}{99}

\bibitem{Cisneros:1970nq}
  A.~Cisneros,
  Astrophys.\ Space Sci.\  {\bf 10} (1971) 87.
\bibitem{Okun:1986hi}   L.~B.~Okun, M.~B.~Voloshin and M.~I.~Vysotsky,
  Sov.\ J.\ Nucl.\ Phys.\  {\bf 44} (1986) 440
  [Yad.\ Fiz.\  {\bf 44} (1986) 677].
\bibitem{Cattadori}
C.~M.~Cattadori, talk at XXI International Conference on Neutrino Physics and
Astrophysics, Neutrino 2004, http://neutrino2004.in2p3.fr/.
\bibitem{Chauhan:2004sf}
  B.~C.~Chauhan and J.~Pulido,
  JHEP {\bf 0406} (2004) 008
  [arXiv:hep-ph/0402194].
\bibitem{Chauhan:2005pn}
  B.~C.~Chauhan, J.~Pulido and R.~S.~Raghavan,
  JHEP {\bf 0507} (2005) 051
  [arXiv:hep-ph/0504069].

\end{thebibliography}
\end{document}